\documentclass[a4paper,12pt]{article}
\usepackage{amssymb,amsmath}
\usepackage{a4wide}
\usepackage{epsfig}
\usepackage{graphics}

\newcommand{\RR}{{\mathbb{R}}}
\newcommand{\CC}{{\mathbb{C}}}
\newcommand{\ZZ}{{\mathbb{Z}}}
\newcommand{\pa}{\partial}
\newcommand{\ii}{{\mathrm{i}}}
\newcommand{\ee}{{\mathrm{e}}}
\renewcommand{\sb}{\bar{s}}
\newcommand{\half}{{\scriptstyle\frac{1}{2}}}
\newcommand{\quar}{{\scriptstyle\frac{1}{4}}}
\newcommand{\tr}{\mathop{\mathrm{tr}}\nolimits}
\renewcommand{\Re}{\mathop{\mathrm{Re}}\nolimits}
\renewcommand{\Im}{\mathop{\mathrm{Im}}\nolimits}
\newcommand{\sgn}{\mathop{\mathrm{sgn}}}
\newcommand{\cM}{{\cal M}}
\newcommand{\cS}{{\cal S}}
\newcommand{\ve}{\varepsilon}

\title{Geometry of Periodic Monopoles}
\author{R.\ Maldonado\footnote{email address: rafael.maldonado@durham.ac.uk}
  \,\, and R.\ S.\ Ward\footnote{email address: richard.ward@durham.ac.uk}
  \bigskip
  \\Department of Mathematical Sciences,
  \\Durham University, Durham DH1 3LE}
\date{\today}

\begin{document}



\maketitle

\begin{abstract}
\noindent
BPS monopoles on $\RR^2\times S^1$ correspond, via the generalized Nahm
transform, to certain solutions of the Hitchin equations on the cylinder
$\RR\times S^1$. The moduli space $\cM$ of two monopoles with their
centre-of-mass fixed is a 4-dimensional manifold with a natural
hyperk\"ahler metric, and its geodesics correspond to slow-motion
monopole scattering. The purpose of this paper is to study the geometry
of $\cM$ in terms of the Nahm/Hitchin data, {\sl ie.}\ in terms of
structures on $\RR\times S^1$. In particular, we identify the moduli,
derive the asymptotic metric on $\cM$, and discuss several geodesic surfaces
and geodesics on $\cM$. The latter include novel examples of monopole
dynamics.
\end{abstract}



\section{Introduction}

This paper deals with periodic BPS monopoles, namely Yang-Mills-Higgs fields
$(\Phi,A_j)$ on $\RR^2\times S^1$ satisfying the Bogomolny equations.
A useful tool for understanding systems of this type is the generalized Nahm
transform. The best-known Nahm transform describes BPS monopoles on
$\RR^3$ in terms of solutions of a set of ordinary differential equations,
namely the Nahm equations. But this is part of a more general picture (which
also includes the ADHM transform for self-dual Yang-Mills instantons):
a generalized Nahm transform, which may be understood in terms of the
reciprocity between self-dual Yang-Mills equations on dual 4-tori \cite{BB89,S88}.
Suitable rescalings of the tori (and corresponding dimensional reductions of
the self-duality equations) then give, as special cases, the ADHM transform and
the Nahm transform for monopoles on $\RR^3$, as well as several other systems
including the present one. The general scheme suggests that the Nahm transforms
of BPS monopoles on $\RR^2\times S^1$ are solutions of the Hitchin equations
on the cylinder $\RR\times S^1$  (satisfying appropriate boundary conditions),
where the two circles are dual to each other. This was confirmed in \cite{CK01},
where many of the details were worked out.

In this paper, we focus on the case of periodic 2-monopole fields with gauge
group SU(2). Such fields may also be visualized as a pair of infinite monopole
chains.  By contrast with the case of monopoles in $\RR^3$, the system as a
whole is infinitely massive, so its centre-of-mass and overall phase
are parameters which must be kept fixed \cite{CK02}. However, the
{\sl relative} separation and phase of the monopoles are free to vary.
So we focus on the relative moduli space $\cM$ of solutions: this is a
4-dimensional manifold equipped with a natural hyperk\"ahler metric.
The geodesics in $\cM$ are of particular interest, as they correspond to
slow-motion dynamics of the system. Our purpose is to study the geometry
of $\cM$ in the Nahm-transformed picture, {\sl ie.}\ in terms of structures
on the cylinder $\RR\times S^1$.  The dynamics of monopoles in this system
is different from that of the well-known case of monopoles on $\RR^3$,
owing to the periodicity. Although the metric on $\cM$ is not known
explicitly, one can identify some geodesics as fixed-point sets of discrete
symmetries, and this provides examples of such novel dynamics.

For the space $\RR^2\times S^1$ on which the monopole fields $(\Phi,A_j)$ 
live, we use coordinates $(x,y,z)$, with $z$ having period $2\pi$.
So  the gauge potential $A_j$ and the Higgs field $\Phi$ are smooth functions
of  $(x,y,z)$, periodic in $z$, and they satisfy the Bogomolny equations
$2D_j\Phi=-\ve_{jkl}F_{kl}$,
where $F_{kl}$ denotes the gauge field. The monopoles are located,
roughly speaking, at the zeros of $\Phi$. For monopole fields of charge~2, the
boundary behaviour in the non-periodic directions, {\sl ie.}\ as $\rho\to\infty$
where $x+\ii y=\rho\,\ee^{\ii\chi}$, is
\[
 \Phi-(\ii/\pi)(\log\rho/C)\,\sigma_3\to0, \quad A_x\to0, \quad A_y\to0, 
      \quad A_z-(\ii\chi/\pi)\,\sigma_3\to0,
\]
locally in some gauge.
Here $C$ is a positive constant which determines the monopole size, or
more accurately the ratio between the monopole size and the $z$-period.
The $C\to0$ limit corresponds to monopoles on $\RR^3$; the opposite
extreme $C\gg1$, where the monopoles spread out, is discussed in more
detail in \cite{M13}.
Note that the system is not rotationally-symmetric about the $z$-axis:
this is reflected in the boundary condition for $A_z$. It does, however,
admit the discrete symmetry of rotations by $\pi$ about the $z$-axis, which
can be compensated by a periodic gauge transformation.

Such monopole fields (or rather, the subset of centred monopole fields)
correspond, via the Nahm transform \cite{CK01}, to the following
Nahm/Hitchin data on $\RR\times S^1$. 
As coordinates on $\RR\times S^1$, take $r\in\RR$ and $t$ with period~$1$.
Let $s$ denote the complex coordinate $s=r+\ii t$. The fields  on
$\RR\times S^1$ consist of a gauge potential $(a_r,a_t)$ with the gauge
group in this case being SU(2), and a complex Higgs field $\phi$ in the adjoint
representation. The gauge field $f_{rt}$ is simply written as $f$. These fields
satisfy the Hitchin equations
\begin{equation} \label{Hit}
  D_{\sb}\phi = 0, \quad 2f = \ii\, [\phi,\phi^*].
\end{equation}
Here $D_{\bar s}\phi:=\pa_{\bar s}\phi+[a_{\bar s},\phi]$, and $\phi^*$ denotes
the complex conjugate transpose of $\phi$. The field $\phi$ is constrained by
\begin{equation} \label{Detphi}
  \det\phi=C^2(-2\cosh(2\pi s)+K),
\end{equation}
where $K$ is some complex constant, and the boundary condition
is $f\to0$ as $r\to\pm\infty$.

In general, for a periodic $n$-monopole system, the Nahm/Hitchin data
would be $\mathfrak{u}(n)$-valued. In our case, $n=2$. The significance
of the fields $(\phi,a_{\sb})$ being $\mathfrak{su}(2)$-valued, rather than
$\mathfrak{u}(2)$-valued, is that the corresponding two-monopole
system is centred, with its centre-of-mass fixed at the point $(0,0,0)$.
As noted above, we are only interested in the relative separation of the
monopoles, and so we restrict to $\mathfrak{su}(2)$.

Actually, one may equally regard the centre-of-mass as being located at
the point $(0,0,\pi)$, since the distinction between these two possibilities
is ambiguous in view of the $z$-periodicity.
There is a natural map $\tau$ which translates a monopole solution by $\pi$
in the $z$-direction, and which therefore interchages these two centring
points. The corresponding map on $(\phi,a_r,a_t)$ consists of
gauging by an antiperiodic gauge transformation, {\sl ie.}\ by
$\Lambda(r,t)\in{\rm SU(2)}$ with $\Lambda(r,1)=-\Lambda(r,0)$ and
$\Lambda_t(r,1)=-\Lambda_t(r,0)$. This preserves the periodicity of
$(\phi,a_r,a_t)$ as well as the equations (\ref{Hit}) and the
boundary conditions. Clearly the fixed points of
$\tau$ consist precisely of the fields which are 1-monopole chains in
disguise, {\sl ie.}\ periodic 1-monopole solutions taken over two periods.
There are exactly two such solutions, up to gauge-equivalence:
their Nahm/Hitchin data can be written as
\begin{equation} \label{SpecialPoints}
  \phi=C(\ee^{\pi s}\mp\ee^{-\pi s})[\cos(\pi t)\sigma_2+\sin(\pi t)\sigma_3],
  \quad a_r=0, \quad a_t=(\ii\pi/2)\sigma_1,
\end{equation}
and they have $K=\pm2$ respectively.

No other solutions of (\ref{Hit}), with these boundary conditions, are
known explicitly. One way of solving the equations numerically is by
minimizing an ``energy'' functional, and we implemented this in order
to get an idea of what the fields look like. Briefly, the details are as
follows. Define
\[
  E_L = \int_{-L}^L \, \int_0^1 \, {\cal E}\,dt\,dr, \mbox{ where }
   {\cal E}=|D_1\phi|^2 + |D_2\phi|^2 + |f|^2 + \quar|[\phi,\phi^*]|^2.
\]
The $r$-cutoff $L$ has to be finite for the integral to converge, but in
practice it does not have to be large, since the solutions are well localized.
Then there is a Bogomolny-type bound on $E_L$, and this bound is saturated
if and only if the Hitchin equations (\ref{Hit}) are satisfied.
So minimizing $E_L$ numerically gives a solution.

In the next section, we identify the four moduli in terms of the fields
$(\phi,a_{\sb})$, at least in the asymptotic region of the moduli space.
Then in Section~\ref{sec3}, we derive the asymptotic metric on $\cM$ by direct
calculation, and see that it agrees with the metric previously derived
by considering the forces between monopoles \cite{CK02}. 
Section~\ref{sec4} describes various geodesic surfaces in $\cM$, and this is
followed by a discussion of geodesics and the associated monopole
dynamics in Section~\ref{sec5}.


\section{The moduli}\label{sec2}
The moduli space $\cM$ is 4-dimensional, and the aim here is to describe
the four moduli in terms of the Nahm/Hitchin data. Two of the moduli are the
real and imaginary parts of the complex number $K$ appearing in the
constraint (\ref{Detphi}). We shall describe the
remaining two moduli in the asymptotic region of $\cM$, which is where
$|K|\gg1$; this corresponds to the monopoles being widely-separated.
Numerical solutions, obtained as outlined above,
indicate that the data $(\phi,a_{\sb})$ then resemble two well separated lumps
on the cylinder, located at the zeros $\pm s_0$ of $\det\phi$.
By this we mean that the gauge field
$f$ is close to zero except at these two points, and the peaks at
$s=\pm s_0$ become more concentrated as $|K|$ increases. In particular,
since $f\approx0$ in the central region $r\approx0$, it makes sense
to consider the $t$-holonomy there, and we define an angle $\theta$ by
\begin{equation} \label{theta}
  U_0={\cal P}\exp\left(\int_0^1 a_t(0,t)\,dt\right),
    \quad 2\cos\theta = \tr U_0.
\end{equation}
The two monopoles are located \cite{CK01, M13, HW09} on
$x+\ii y=\pm C\sqrt{-K}$, and $\theta$ determines their $z$-offset:
the $z$-coordinates of the monopoles are
$z=\pm\theta$ respectively, as we shall see in the next section.
The sign ambiguity in the square root reflects the indistinguishability of the
monopoles.

The equation (\ref{theta}) only determines $\theta$ up to sign,
but we may remove this ambiguity by using the Higgs field $\phi$, and regard
$\theta\in(-\pi,\pi]$ as periodic with period $2\pi$. In fact, $\theta$ is
really a local coordinate on a $\ZZ_2$-twisted circle bundle over the
asymptotic $K$-space, and the definition of its sign is a matter of choice.
Here is one particular scheme, using the value $\phi_0$ of $\phi$
at $r=t=0$. The quantity
$\xi=-\ii\tr(U_0 \phi_0)$ is a gauge-invariant complex number.
If $\theta\neq0,\pi$, then $\xi$ is non-zero, and it follows from
$\det\phi_0=C^2(K-2)\approx C^2K$ that
\begin{itemize}
\item if $\Re(K)\geq0$, then $\Im(\xi)\neq0$, and we can define
       $\sgn(\theta)=\sgn(\Im(\xi))$;
\item if $\Re(K)<0$, then $\Re(\xi)\neq0$, and we can define
       $\sgn(\theta)=\sgn(\Re(\xi))$.
\end{itemize}
If we write $K=|K|\ee^{2\pi\ii\eta}$, then $\theta$ changes sign as $\eta$ goes
from $0$ to $1$, {\sl ie}\ as one goes around a loop in the asymptotic
$K$-space. In fact, with this particular scheme, the jump occurs as $\eta$
crosses $1/4$.

Finally, let us turn to the fourth modulus, $\omega$, which corresponds to a
relative phase between the two monopoles. It also correponds to a relative
phase between the two lumps on the cylinder. Define
$f_{\pm}=f(\pm s_0)\in\mathfrak{su}(2)$: in other words, the $f_{\pm}$
are the directions in the Lie algebra of the gauge field at the two peaks.
Then we define $\omega$ to be the angle between $f_+$ and $\hat{f}_-$,
where $\hat{f}_-$ is obtained by parallel-propagating $f_-$ from $-s_0$ to
$s_0$ using $\partial_\gamma f=-[a_\gamma,f]$, where $\gamma$ is a path 
between $-s_0$ and $s_0$.  For large $|K|$ this formulation is path independent 
(up to winding round the cylinder) as the field strength vanishes between 
the peaks.  This only defines $\omega$ up to sign, but the ambiguity may be 
resolved as before,this time using the quantity $\xi=\ii\tr(f_+ f_- \phi_0)$. 
So $\omega\in(-\pi,\pi]$ has period $2\pi$.

Since $f\to0$ as $r\to\pm\infty$, it is also natural to consider the holonomies
at infinity, namely
\[
   U_{\pm} = \lim_{r\to\pm\infty}{\cal P}\exp\left(\int_0^1 a_t(r,t)\,dt\right).
\]
It follows from the equations and boundary conditions that $\tr U_{\pm}=0$,
so the $U_{\pm}$ individually contain no gauge-invariant information. But
the angle between them does: for example, let $\hat{U}_-$ be the element
of SU(2) obtained by parallel-propagating $U_-$ along $t=0$ from $r=-\infty$
to $r=\infty$, and define $\tilde\omega$ by
$2\cos\tilde\omega = \tr\left(U_+ \hat{U}_-^*\right)$. This quantity $\tilde\omega$
is related to $\omega$ by $\tilde\omega-\omega=\pi$ (modulo integer multiples
of $2\pi$). By contrast with $\omega$, the definition of $\tilde\omega$ is valid
throughout $\cM$, and not just in the asymptotic region. In particular,
for the special solutions (\ref{SpecialPoints}), which are not in the asymptotic
region, we can compute $U_{\pm}$ exactly, and this shows that the fields
(\ref{SpecialPoints}) have $\tilde\omega=0$.


\section{The asymptotic metric}\label{sec3}
The natural hyperk\"ahler metric on $\cM$ is believed to have no continuous
symmetries, and it is not known explicitly. The asymptotic metric, however,
has a fairly simple form. It was derived in \cite{CK02} by studying the
effective Lagrangian of the 2-monopole system, in other words the forces
between well separated monopoles. In this section, we see that
this asymptotic metric can be calculated directly in terms of the
Nahm/Hitchin data. In particular, this shows us how to identify the
moduli of the previous section with the ``monopole-based" moduli
used in \cite{CK02}.

Let us think of a tangent vector in $\cM$, at the point corresponding to
the solution $(\phi,a_{\sb})$, as a perturbation $(\delta\phi,\delta a_{\sb})$
which preserves the equations (\ref{Hit}), and also satisfies the condition
\begin{equation} \label{GaugeCondition}
4\left\{D_s(\delta a_{\sb})+D_{\sb}(\delta a_s)\right\}=
               [\phi,\delta\phi^*]+[\phi^*,\delta\phi]
\end{equation}
for the perturbation to be orthogonal to the gauge orbits at $(\phi,a_{\sb})$.
Here $\delta a_{\sb}=\half(\delta a_r+\ii\delta a_t)$ and
$\delta a_s=\half(\delta a_r-\ii\delta a_t)=-(\delta a_{\sb})^*$.
The combined equations on $(\delta\phi,\delta a_{\sb})$ are equivalent to
the pair
\begin{equation} \label{Perturbation}
  D_{\sb}(\delta\phi)=[\phi,\delta a_{\sb}], \quad
  [\phi,\delta\phi^*]=4D_{\sb}(\delta a_s).
\end{equation}
In addition to these differential equations (\ref{Perturbation}), one also
needs boundary conditions $\delta\phi\to0$, $\delta a_{\sb}\to0$ as $r\to\infty$,
and the constraint $\tr(\phi\,\delta\phi)=\mbox{constant}$. The norm-squared
of the vector $V=(\delta\phi,\delta a_{\sb})$ is then defined to be
\begin{equation} \label{MetricIntegral}
 \|V\|^2 =\frac{1}{2}\Re\int\tr\left[(\delta\phi)(\delta\phi)^*+
   4(\delta a_{\sb})(\delta a_{\sb})^*\right]\,dr\,dt,
\end{equation}
and this gives the metric on $\cM$.

Note that if $V_1=(\delta_1\phi,\delta_1 a_{\sb})$ is one solution of
(\ref{Perturbation}), then so are each of
\begin{eqnarray*} 
V_2 &=& (\delta_2\phi,\delta_2 a_{\sb})=(\ii\delta_1\phi,\ii\delta_1 a_{\sb}),\\
V_3 &=& (\delta_3\phi,\delta_3 a_{\sb})=(2\delta_1 a_s,\half\delta_1\phi^*),\\
V_4 &=& (\delta_4\phi,\delta_4 a_{\sb})=(2\ii\delta_1 a_s,\half\ii\delta_1\phi^*).
\end{eqnarray*}
Furthermore, these four vectors are orthogonal with the same norm; in other
words, $\langle V_a,V_b\rangle=p^2\delta_{ab}$ for some real constant $p$.
Suppose that $V_1$ corresponds to the increments
$(\delta_1 K_r, \delta_1 K_i, \delta_1\theta, \delta_1\omega)$
in the moduli, where $K=K_r+\ii K_i$ are the real and imaginary parts of
$K$; and similarly for the other $V_a$. Define the $4\times4$ matrix $Q$ by
\[
  Q=p^{-1}\left[\begin{array}{cccc}
       \delta_1 K_r & \delta_2 K_r & \delta_3 K_r &\delta_4 K_r\\
       \delta_1 K_i & \delta_2 K_i & \delta_3 K_i &\delta_4 K_i\\
   \delta_1\theta & \delta_2\theta & \delta_3\theta & \delta_4\theta\\
   \delta_1\omega & \delta_2\omega & \delta_3\omega & \delta_4\omega
              \end{array}\right].
\]
Then the coefficients of the metric on $\cM$, with respect to the local
coordinates $(K_r,K_i,\theta,\omega)$, are the entries in the matrix
$g=(Q Q^t)^{-1}$.

In the asymptotic region $|K|\gg1$, there is a crude but effective
approximate solution $(\phi,a_{\sb})$, namely
\begin{equation} \label{SingularSoln}
   \phi=C\sqrt{H}\sigma_3, \quad a_t=\ii\theta\sigma_3, \quad a_r=0,
\end{equation}
where $H=2\cosh(2\pi s)-K$. There are branch cuts along $(r\geq r_0,\,t=t_0)$ and 
$(r\leq-r_0,\,t=-t_0)$, on which $a_t\neq0$, and (\ref{SingularSoln}) is a solution everywhere 
except at the two singular points $s=\pm s_0$. Numerics indicate that it is a good
approximation, for large $|K|$, to the actual smooth solutions, except
very close to the singular points. The gauge field consists, in effect,
of delta-functions at the singular points, and we complete the
approximate description of the field by simply assigning elements
$f_{\pm}$ of $\mathfrak{su}(2)$, orthogonal to $\sigma_3$, to these
two points. The moduli $K_r$, $K_i$ and $\theta$ appear explicitly in
(\ref{SingularSoln}), and $\omega$ is the angle between $f_+$ and
$f_-$.

To keep things simple in what follows, let us restrict to the case
$\Re(K)>0$. Since $\theta$ and $\omega$ are twisted rather than global
coordinates, one obtains the complete picture by doing the case
$\Re(K)<0$ as well, and then patching things together in an appropriate
way.

Let $V_1$ be the perturbation $\delta_1\phi=\half\ve h\sigma_3$,
$\delta_1 a_{\sb}=0$, where $h=h(s)=H^{-1/2}$, and $\ve$ is a small
parameter. This satisfies
(\ref{Perturbation}) and the associated boundary conditions, except at
the singularities. Note that $\|V_1\|^2=p^2=\ve^2 I$, where
\begin{equation} \label{I}
  I=\frac{1}{4} \int\frac{dr\,dt}{|H|}
    =\frac{1}{4} \int\frac{dr\,dt}{|2\cosh(2\pi s)-K|}\,.
\end{equation}
The corresponding variations in the moduli are $\delta_1 K_r=\ve C^{-1}$
and $\delta_1 K_i=\delta_1\theta=\delta_1\omega=0$.
Following the pattern described above, the next perturbation $V_2$ is
$\delta_2\phi=\half\ii\ve h\sigma_3$, $\delta_2 a_{\sb}=0$, and this has
$\delta_2 K_i=\ve C^{-1}$, $\delta_2 K_r=\delta_2\theta=\delta_2\omega=0$.

For $V_3$ we get $\delta_3\phi=0$, whence $\delta_3 K_r=\delta_3 K_i=0$;
and $\delta_3 a_{\sb}=\quar\ve \bar{h}\sigma_3$. Thus
$\delta_3 a_r=-\half\ii\ve\Im(h)\sigma_3$ and 
$\delta_3 a_t=-\half\ii\ve\Re(h)\sigma_3$, from which one directly computes
$\delta_3\theta=-\half\ve\Re(h_0)$, where
\[
   h_0 = \int_0^1 h(0,t)\,dt \approx 1/\sqrt{-K}.
\]
The perturbation $V_3$ does not affect $f_\pm$ (this has been checked with 
numerical examples), so a variation of $\omega$ arises only from the change 
of $a$ along the path between $-s_0$ and $s_0$.  We thus obtain 
$\delta_3\omega = -2\ve[\Im(J)+\Re(L)]$, where
\[
    J=\int_0^{r_0} h(r,t_0)\,dr, \quad
     L=\int_0^\eta h(0,t)\,dt\approx h_0\eta
\]
(recall that $\eta\in[0,1)$ is defined by $K=|K|\ee^{2\pi\ii\eta}$).
Similarly, for $V_4$ we have $\delta_4\phi=0$ and
$\delta_4 a_{\sb}=\quar\ii\ve \bar{h}\sigma_3$; and the corresponding
variations in the moduli are $\delta_4 K_r=\delta_4 K_i=0$,
$\delta_4\theta=-\half\ve\Im(h_0)$ and
$\delta_4\omega=2\ve[\Re(J)-\Im(L)]$.   Note that the `knock-on' effect of 
one perturbation on another has been ignored, and indeed for large $|K|$ such terms 
only make relatively small contributions to the metric.

The next step is to compute the leading terms in the integrals $I$ and
$J$ for $|K|\gg1$. These are obtained from the approximation
$1/H(r,t)\approx-1/K$ for $0\leq r<r_0$, $1/H(r,t)\approx0$ for $r>r_0$, 
which gives
\[
   I\sim\frac{\log|K|}{4\pi|K|}, \quad J\sim\frac{\log|K|}{2\pi\sqrt{-K}}
\]
as $|K|\to\infty$.
Then it is straightforward to calculate the asymptotic metric as
described above, via the matrix $Q$, and we get
\begin{equation} \label{AsympMetric}
  ds^2 = \frac{\log|K|}{4\pi|K|}\left(C^2|dK|^2+4|K|d\theta^2\right)
         + \frac{\pi}{4\log|K|}\left(d\omega-4\eta\, d\theta\right)^2.
\end{equation}

Now the asymptotic metric $ds_{CK}^2$ of \cite{CK02}, which was
computed by considering the forces between monopoles, is given by
\begin{equation} \label{CKmetric}
\frac{1}{4\pi} ds_{CK}^2 =4U(dx^2+dy^2+dz^2)+U^{-1}(d\nu + g\,dz)^2,
\end{equation}
where $(x,y,z)$ is the location of one of the monopoles relative to the
centre of mass, and $\nu$ is a relative phase with period $\pi$. The functions
$U$ and $g$ are defined by
\[
   U=\frac{1}{\pi}\log\rho, \quad g=2(\chi-\chi_0),
\]
where $x+\ii y=\rho\,\ee^{\ii\pi(\chi-\chi_0)}$, with $\chi_0$ being some
constant. We already know \cite{CK01, M13} that $x+\ii y=C\sqrt{-K}$;
and it is straightforward to check that the metrics (\ref{AsympMetric})
and (\ref{CKmetric}) agree, up to an overall factor of $8\pi$, if we make
the identification $\theta=z$ and $\omega=-2\nu$. In particular, therefore,
$2\theta$ is the $z$-offset of the monopoles.


\section{Geodesic surfaces}\label{sec4}

The geometry of the moduli space $\cM$, and in particular its geodesics,
correspond to the dynamics of monopole systems in situations where radiative
losses are small, and in particular when speeds are small \cite{M82,MS04}.
Not knowing the metric explicitly means that we cannot find many geodesics
exactly; but some can be obtained as fixed-point sets
of discrete symmetries of $\cM$. The first step, as in the familiar $\RR^3$
case \cite{AH88}, is to identify geodesic surfaces in $\cM$; we do this by
looking for discrete symmetries of the system (\ref{Hit}, \ref{Detphi}).
The most obvious symmetry is
\begin{equation} \label{phi_to-phi}
   \phi\mapsto-\phi, \quad a_{\sb}\mapsto a_{\sb}.
\end{equation}
In the monopole picture, (\ref{phi_to-phi}) corresponds to rotation
by $\pi$ about the $z$-axis. Let $\cS$ denote the fixed-point set of
(\ref{phi_to-phi}). Since $K$ is preserved, (\ref{phi_to-phi}) acts only
on the other two
moduli $\theta$ and $\tilde\omega$ (for this section, it is more convenient
to use $\tilde\omega$ than $\omega$). From the discussion of signs in
Section~\ref{sec2}, it is clear that the effect of (\ref{phi_to-phi}) is
$\theta\mapsto-\theta$ and $\tilde\omega\mapsto-\tilde\omega$.
Thus asymptotically, $\cS$ has four disconnected components,
corresponding to $\theta,\tilde\omega\in\{0,\pi\}$. 
In this asymptotic regime, we see a pair of monopoles, located at
the points $x+\ii y=\pm C\sqrt{-K}$, $z=\pm\theta$, and with their
phases either aligned or anti-aligned depending on $\tilde\omega$.

The question now is what $\cS$ looks like globally, not just asymptotically.
As described in \cite{HW09}, the solutions belonging to $\cS$ take a
simplified form: there exists a gauge in which
\[
  a_{\sb}=h\sigma_1,\quad \phi=\half(f+g)\sigma_2+\half\ii(f-g)\sigma_3,
\]
where $f$, $g$ and $h$ are complex-valued functions.
The constraint on $\det\phi$ is $fg=C^2H=C^2[2\cosh(2\pi s)-K]$,
and the Hitchin equations become
\begin{equation} \label{HWeqns}
  \Delta\log|f|=|f|^2-C^4|H/f|^2,
\end{equation}
together with $2h=-\pa_{\sb}\log f$, where $\Delta=4\pa_s\pa_{\sb}$
is the Laplacian. The boundary condition is $|f^2/H|\to C^2$ as
$r\to\pm\infty$, and the solutions have the symmetry
\begin{equation} \label{f-symmetry}
  |f(-r,1-t)|=|f(r,t)| \mbox{ for all $r,t$}.
\end{equation}
The remaining gauge freedom consists of 
\[
  \{f\mapsto g, \, g\mapsto f, \, h\mapsto-h\} \mbox{ and }
  \{f\mapsto\lambda^{-2}f, \, g\mapsto\lambda^{2}g, 
      \, h\mapsto h+\pa_{\sb}\lambda\},
\]
where $\lambda(r,t)$ is periodic and of unit modulus.

Taking account of this gauge freedom, it is easy to see that there
are four classes of solution
of (\ref{HWeqns}): $f$~could have one zero or none; and
$\lim_{r\to\infty}\Im\log f(r,t)=2\pi nt$, where $n$ is either $0$ or $1$.
As we shall see below, these four possibilities correspond to the four
asymptotic components of $\cS$. Let us begin by studying one of these
cases in detail, namely where $f$ has no zeros and $n=0$. In particular,
this means that $f$ has the form
\begin{equation} \label{formA}
  f=\ee^{\psi/2},
\end{equation}
where $\psi$ is a complex-valued periodic function.
If we write $\psi=\alpha+\ii\beta$ for the real and imaginary parts
of $\psi$, then $a_t=\quar\ii(\alpha_r-\beta_t)\sigma_1$, and therefore the
holonomy at $r$ is
\[
  U_r = \exp\left[\frac{\ii}{4}\int_0^1\alpha_r(r,t)\,dt \,\sigma_1\right].
\]
The boundary condition says that $\alpha(r,t)\sim\pm2\pi r$ as
$r\to\pm\infty$, so the holonomies at the two ends are
$U_{\pm}=\pm\ii\sigma_1$, and hence $\tilde\omega=\pi$. To compute
$\theta$, note that the symmetry (\ref{f-symmetry}) implies
$\alpha_r(0,1-t)=-\alpha_r(0,t)$; it follows that $U_0=1$ and so
$\theta=0$.

For $f$ of the form (\ref{formA}), the equation (\ref{HWeqns})
becomes
\begin{equation} \label{alpha-eqn}
  \Delta\alpha=2\left(\ee^{\alpha}-C^4|H|^2\ee^{-\alpha}\right),
\end{equation}
which has a unique solution for each value of $K\in\CC$.
So this case gives one of the components $\cS_-$ of $\cS$, namely
the one corresponding to $\theta=0$, $\tilde\omega=\pi$; and $\cS_-$
is diffeomorphic to $\CC$. With its natural metric (the
restriction of the metric on $\cM$), the surface $\cS_-$
is a deformed version of the Atiyah-Hitchin cone \cite{AH88,MS04},
having no continuous symmetries (unlike the Atiyah-Hitchin cone
itself, which is rotationally-symmetric). The metric on $\cS_-$ may
be calculated numerically, and this is instructive in that it shows
the effect of varying the parameter $C$. The procedure is as follows.

\begin{figure}
\begin{center}
\includegraphics[scale=1.0]{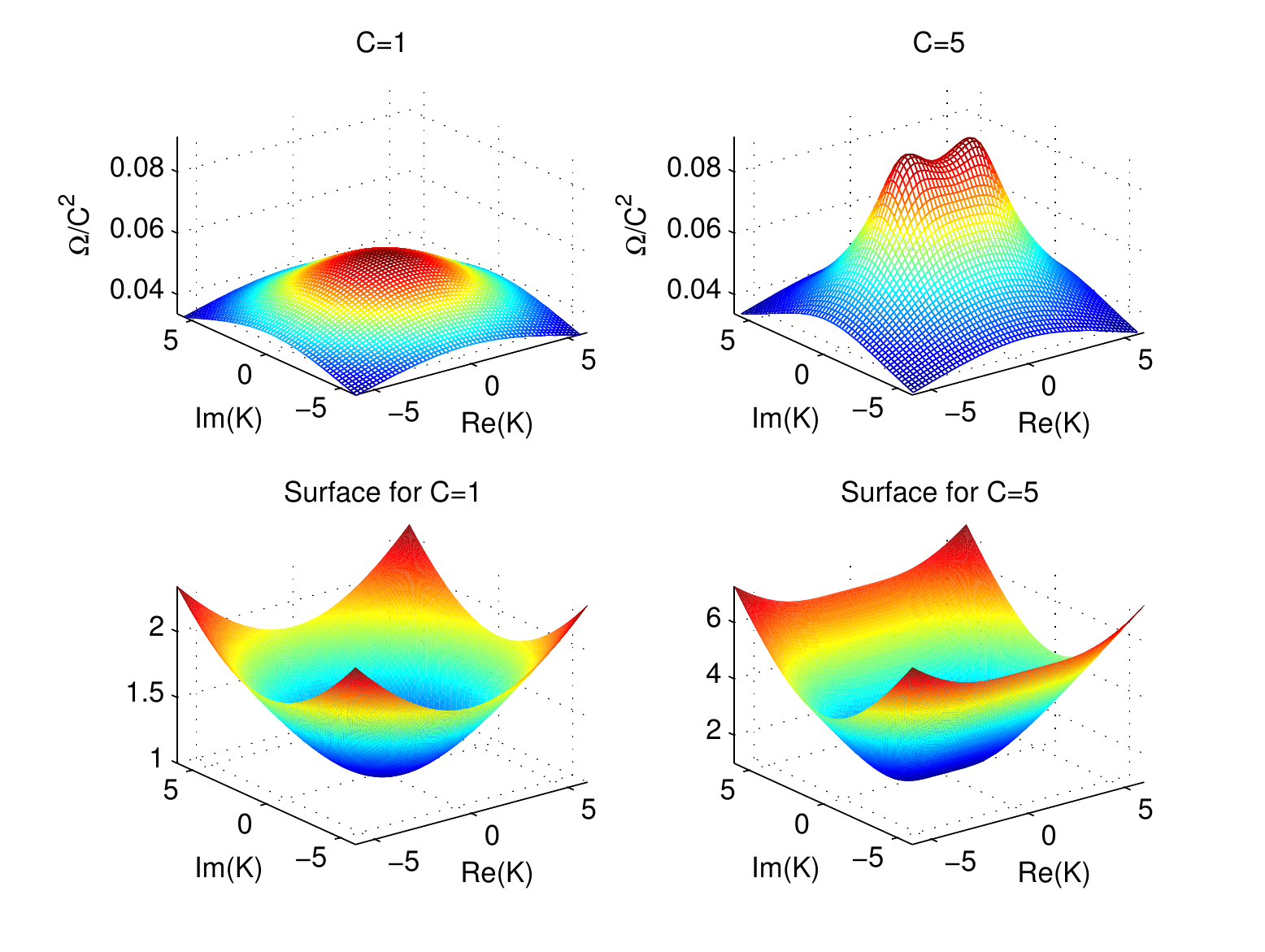}
\caption{The metric factor $\Omega/C^2$ on the geodesic surface $\cS_-$,
         and sketches of the surface, for $C=1$ and $C=5$.} \label{fig1}
\end{center}
\end{figure}

Given complex numbers $K$ and $\delta K$, with $\delta K$ small,
we solve (\ref{alpha-eqn}) numerically for $K$ and for $K'=K+\delta K$,
giving real functions $\alpha$ and $\alpha'$ respectively. This may be
done by minimizing an appropriate functional of $\alpha$, as described
in \cite{HW09}. We take $\beta=0$, so $\psi=\alpha$ is real-valued:
this is just a gauge choice. However, setting $\beta'=0$ as well
leads to a perturbation $(\delta\phi,\delta a_{\sb})$ which does
not satisfy the gauge-orthogonality condition (\ref{GaugeCondition}).
So we need to use $\delta\psi=\alpha'-\alpha+\ii\,\delta\beta$, where
$\delta\beta$ is determined by the requirement that $\delta\psi$ be
orthogonal to the gauge orbits at $\psi=\alpha$. This is just a
linear equation for $\delta\beta$, having a unique solution, and is
straightforward to solve numerically. The final step then uses
(\ref{MetricIntegral}) to evaluate $\|\delta K\|^2$, and hence
gives the metric on $\cS_-$. This metric has the form
$ds^2 = \Omega(K)\,|dK|^2$, and we know from (\ref{AsympMetric}) that
\[
  \Omega(K) \sim \frac{C^2}{4\pi|K|} \log|K| \mbox{ as $|K|\to\infty$}.
\]
The upper plots in Figure~\ref{fig1} show $\Omega(K)/C^2$ on $|K|<6$,
for $C=1$ and $C=5$ respectively. The corresponding lower plots are 
rough sketches of the surface, obtained by computing the Gaussian
curvature numerically from $\Omega$ and then finding an embedded
surface in $\RR^3$ with that curvature.
We see that for $C=1$, the surface has approximate rotational symmetry,
as one would expect since it should approach the Atiyah-Hitchin cone
as $C\to0$. For larger $C$, however, the lack of symmetry becomes
apparent, with the cone becoming stretched in the $K_r$-direction.

We next consider the component of $\cS$ corresponding to the
case where $f$ has no zeros and $n=1$. The corresponding geodesic surface
$\cS_+$ is isometric to $\cS_-$: in fact, the isometry is the map $\tau$.
The action of $\tau$ amounts to gauging by the
antiperiodic transformation $\Lambda=\exp(\ii\pi t\sigma_3)$, and the effect
of this on the moduli is $K\mapsto K$, $\theta\mapsto\theta+\pi$, 
$\tilde\omega\mapsto\tilde\omega$.
So in particular, $\cS_+$ has $\theta=\tilde\omega=\pi$.

For the two remaining cases, $f$ has a zero, which has to be one of
the zeros of $H$. Then $f$ has either of the two forms
\begin{eqnarray} 
f &=& C\ee^{\psi/2}\mu\,\ee^{\ii\pi t}, \label{formC}\\
f &=& C\ee^{\psi/2}\mu\,\ee^{-\ii\pi t}, \label{formD}
\end{eqnarray}
where
\begin{equation} \label{mu}
  \mu=\ee^{\pi s}-W\,\ee^{-\pi s}, \quad W=(K+\sqrt{K^2-4})/2,
\end{equation}
and we take the branch of the square root such that $|W|>1$.
The boundary condition is $\Re(\psi)\to0$ as $r\to\infty$, 
$\Re(\psi)\to-2\log|W|$ as $r\to-\infty$. These classes
(\ref{formC}, \ref{formD}) both have $\tilde\omega=0$, and they
have $\theta=0,\pi$ respectively. They are interchanged by the map~$\tau$.
However, each class contains the two special solutions (\ref{SpecialPoints}),
which are the fixed points of~$\tau$. So in fact we get a single
component $\cS_0$ of $\cS$, consisting of two copies of the $K$-plane
branched over the points $K=\pm2$. In effect, interchanging the two
branches of the square root in (\ref{mu}) interchanges the forms
(\ref{formC}) and (\ref{formD}). The single surface $\cS_0$ has two
asymptotic regions, each of which is cone-like: so the picture may be
described as a double-trumpet, by contrast with the Atiyah-Hitchin
trumpet of the $\RR^3$ case \cite{AH88}. The metric on $\cS_0$ has
no continuous symmetries, but has an approximate rotational symmetry
(about the axis of the trumpet) for small $C$.

The second expression in (\ref{mu}) is just the usual conformal mapping
$K=W+W^{-1}$, and this gives us a global coordinate $W\in\CC^*$ on
$\cS_0$. Given $W$, we take the field to be determined by (\ref{formD}).
If $|W|\gg1$ it lies on the sheet $\theta=\pi$, while if $|W|\ll1$
it lies on the sheet $\theta=0$. The $K$-plane is cut on the line segment
$-2\le K\le2$, which corresponds to $|W|=1$, and crossing this curve
goes from one sheet to the other. In the next section, we shall describe
geodesics which cross sheets in this way.

\begin{figure}
\begin{center}
\includegraphics[width=10cm]{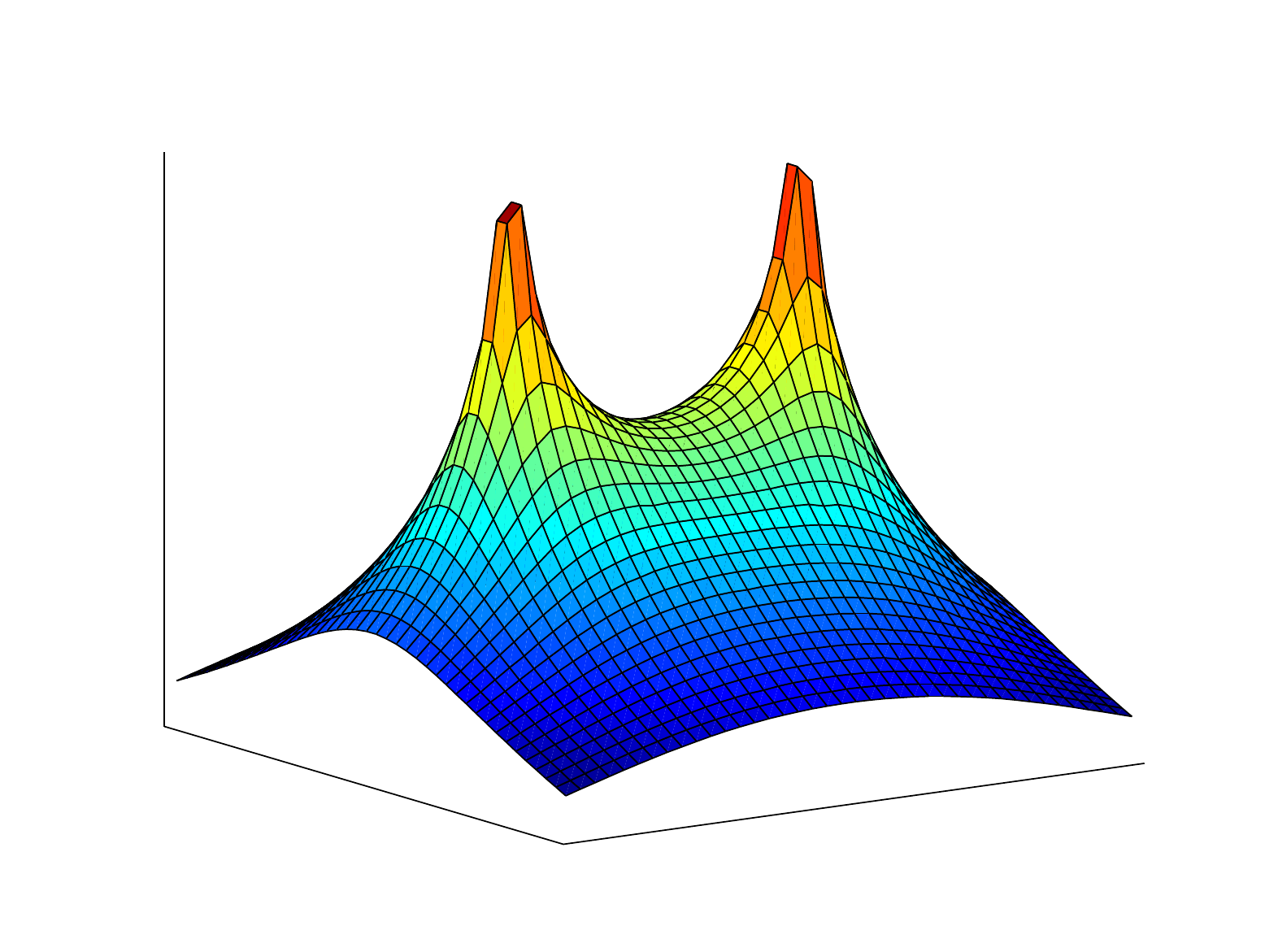}
\caption{Plot of the large-$C$ conformal factor $I(K)$.
   The peaks are at $K=\pm2$.}\label{fig2}
\end{center}
\end{figure}

Finally, we remark on the large-$C$ behaviour of the metric.
In \cite{M13} it was suggested that when $C\gg1$, {\sl ie.}\  when the
monopoles are large compared to the $z$-period, the only relevant
modulus is $K$; and the metric on the $K$-plane was computed by
using an approximation to the monopole fields which is valid in this limit.
In fact, we can also compute this limiting metric in the Nahm-transformed
picture. The crucial observation is that in the large-$C$ regime, the
fields are well approximated by the singular solution (\ref{SingularSoln}),
except at its singularities. In other words, the large-$|K|$ and large-$C$
approximations are the same. It follows that the large-$C$ metric on
each component of $\cS$ is $ds^2=C^2 I |dK|^2$, where $I(K)$ is
defined in (\ref{I}). In Section~\ref{sec3} we only used $I(K)$ for $|K|$ large,
but in this context we need it for all $K\in\CC$. The integral (\ref{I})
converges for all $K$ except the two special values $K=\pm2$, and
its value is plotted in Figure~\ref{fig2}, as a function of $K\in\CC$. This
picture should be compared with the plots of $\Omega/C^2$ in
Figure~\ref{fig1} for $C=1$ and $C=5$: the function $I(K)$ appears to
be the $C\to\infty$ limit of $\Omega/C^2$. A more extensive
numerical investigation of  $\Omega$, for a wider range of $C$,
bears this out. Note also that in this limit, the surface $\cS_0$
resembles two copies of Figure~\ref{fig2}, branched between the singularities.


\section{Geodesics and monopole scattering}\label{sec5}

Our aim in this section is to identify geodesics on $\cS_{\pm}$ and $\cS_0$,
and to interpret these in terms of 2-monopole trajectories.
One could construct such geodesics numerically, for example 
by using the numerically-derived metrics on these surfaces; such
a construction was implemented in \cite{M13} in the large-$C$
limit. But here we will do something more analytic,
namely identifying geodesics on $\cS_{\pm}$ and $\cS_0$ as fixed-point sets
of additional symmetries of the system. An example of this type was
presented in \cite{HW09}; we will here give a fuller discussion, revealing
rather more interesting behaviour than was seen before.

The first step is to describe the relevant symmetries of the Hitchin
equations (\ref{Hit}). There are two of them, namely
\begin{equation}\label{Sym1}
  K\mapsto\overline{K},\, \phi(r,t)\mapsto\phi(r,1-t)^*,\,
   a_r(r,t)\mapsto a_r(r,1-t),\, a_t(r,t)\mapsto-a_t(r,1-t),
\end{equation}
\begin{equation}\label{Sym2}
  K\mapsto-\overline{K},\, \phi(r,t)\mapsto\ii\phi(r,\half-t)^*,\,
  a_r(r,t)\mapsto a_r(r,\half-t),\, a_t(r,t)\mapsto-a_t(r,\half-t).
\end{equation}
The fixed-point sets of (\ref{Sym1}) and  (\ref{Sym2}) are geodesic
hypersurfaces in
$\cM$ given by $K\in\RR$ and $K\in\ii\RR$ respectively, and the
intersections of these with the surfaces $\cS_{\pm}$ and $\cS_0$ are
geodesics. For $\cS_{\pm}$, this leads to a picture which is essentially
the same as in the $\RR^3$ case: the geodesics pass over the ``centre''
of the deformed cone, and this corresponds to $90^\circ$ planar
scattering of two monopoles, via a toroidal 2-monopole solution.
The geodesics on the double-trumpet $\cS_0$ are more interesting,
however, and we shall focus on them in what follows.

In terms of the coordinate $W\in\CC^*$ on $\cS_0$, the symmetries
(\ref{Sym1}, \ref{Sym2}) lead to five complete geodesics, namely the four
half-axes in the $W$-plane and the unit circle $|W|=1$.
The last of these is a closed geodesic which winds around the waist
of the double-trumpet. The two points $W=\pm1$ on it are the two special
solutions (\ref{SpecialPoints}) representing 1-monopole fields taken over
two periods, the monopoles being located on the $z$-axis at $z=\pm\pi/2$.
\begin{figure}
\begin{center}
\includegraphics[scale=0.7]{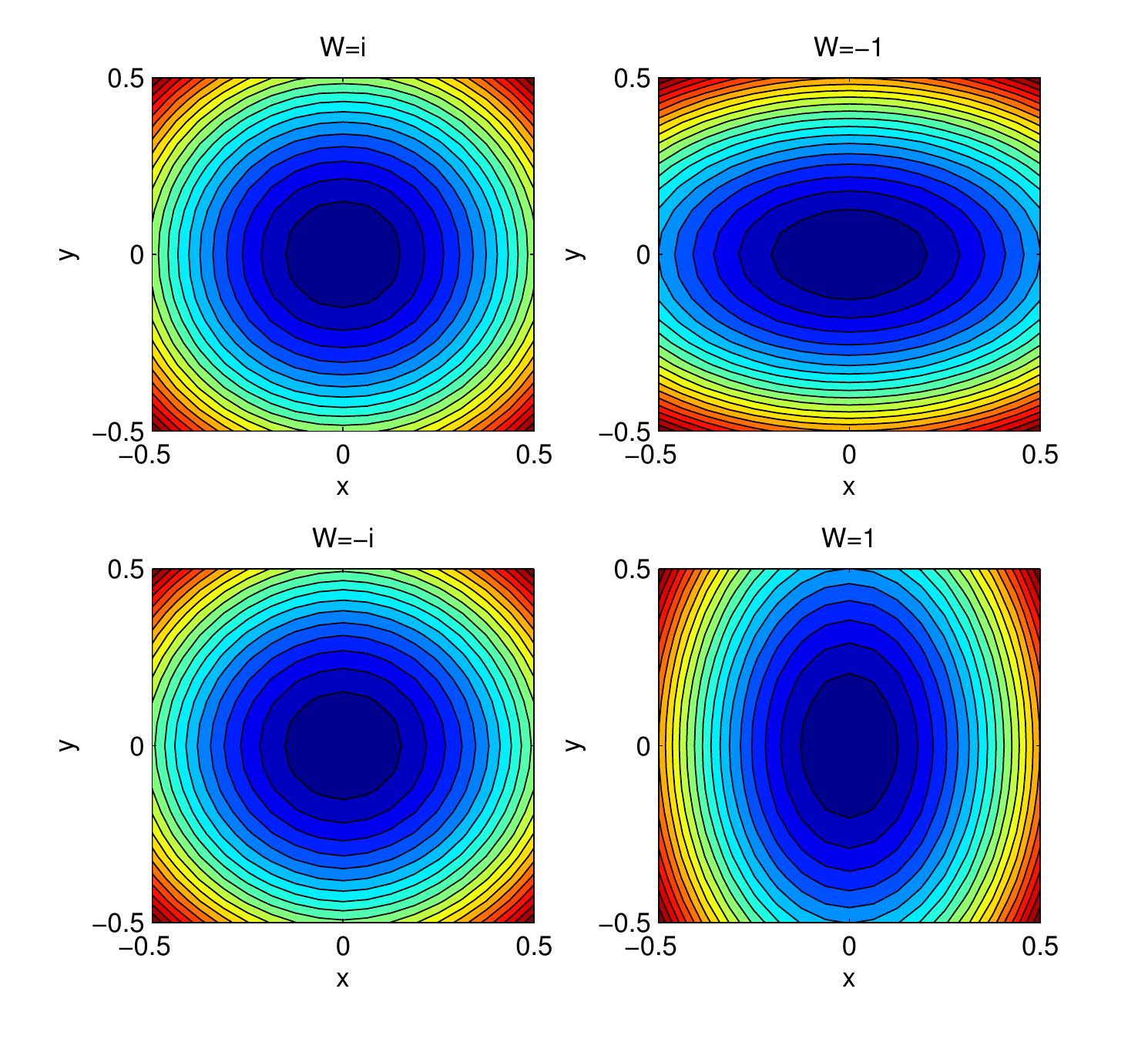}
\caption{Contour plots of $|\Phi|^2$ on $z=\pi/2$,
           for four points on the geodesic $|W|=1$.} \label{fig3}
\end{center}
\end{figure}
In fact, the geodesic consists entirely of monopole pairs located at these
two points on the $z$-axis: the monopoles stay in the same position and
simply oscillate in shape. This is illustrated in Figure~\ref{fig3}, which was
generated by numerical solution of the Hitchin equations (\ref{HWeqns})
for a range of $W$-values, followed by numerical implementation of the
inverse Nahm transform. The parameter $C$ was taken to  have the value
$C=1$. Each of the plots in Figure~\ref{fig3} is a contour plot
of $|\Phi|^2$ on the plane $z=\pi/2$, with the Higgs field $\Phi$ having a
zero at the centre $x=y=0$. The cases $W=\pm\ii$ correspond to $K=0$
on each of the two $K$-sheets, and one then has an additional
$x\leftrightarrow y$ symmetry which is absent for the other points on this
bounded trajectory.
The pictures on the plane $z=-\pi/2$ are the same. So we have
a periodic trajectory representing a string of equally-spaced monopoles,
with all their kinetic energy coming from their in-phase shape oscillation.

\begin{figure}
\begin{center}
\includegraphics[scale=0.6]{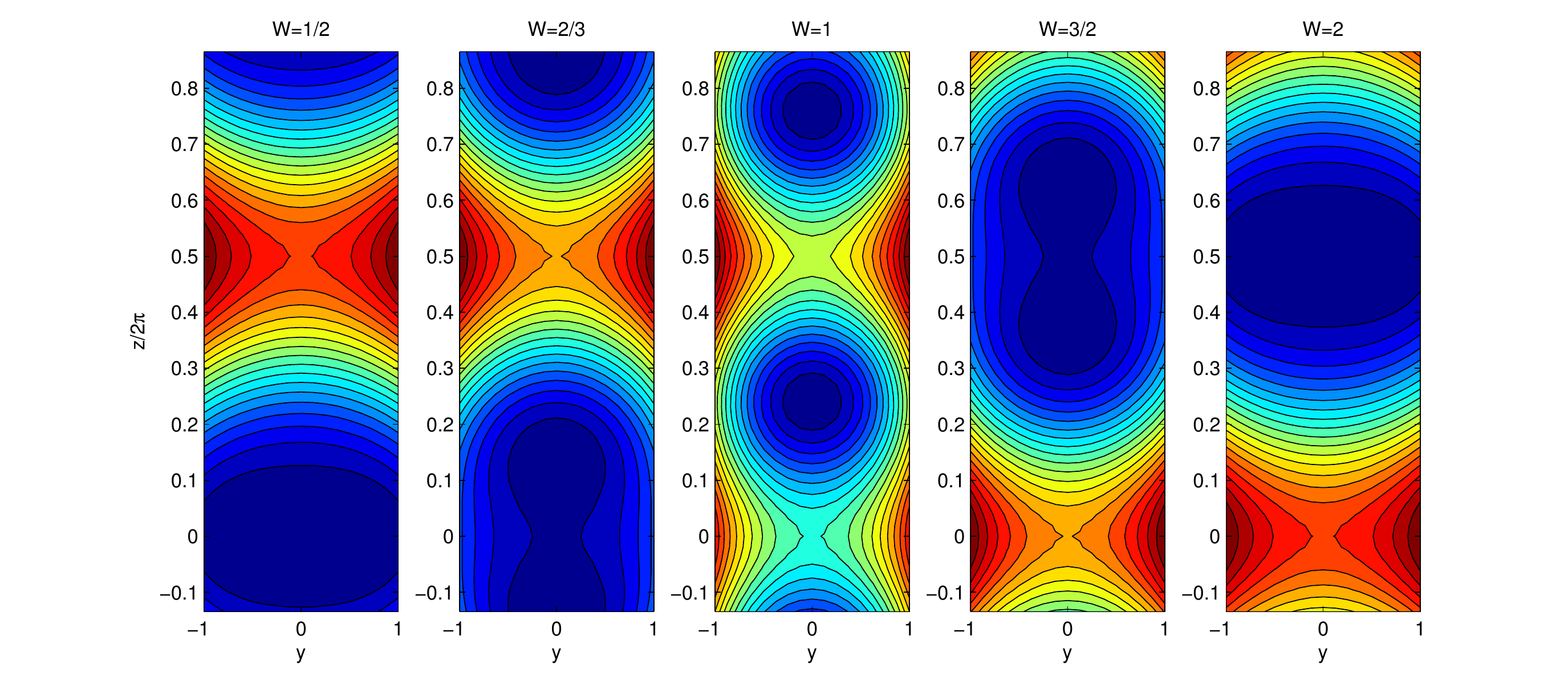}
\caption{Contour plots of $|\Phi|^2$ on $x=0$,
           for five points on the geodesic $W>0$.} \label{fig4}
\end{center}
\end{figure}

The other four geodesics mentioned above are of scattering type, where
two widely-separated monopoles undergo a head-on collision and then
separate again. Let us first describe the $W>0$ case, {\sl ie.}\ $W$ on
the positive real axis. A point with $0<W\ll1$ corresponds to a pair of
monopoles widely-separated on the $y$-axis, in fact at $x=0=z$,
$y=\pm C/\sqrt{W}$.
At the other end of the geodesic, where $W\gg1$, we have monopoles
at $x=0$, $y=\pm C\sqrt{W}$, $z=\pi$. In other words, the monopoles
approach each other along the $y$-axis, collide, and emerge along the
$\pm y$ directions but shifted by half a period in $z$. Using a numerical
Nahm transform for a sequence of real $W$-values near $W=1$ reveals
what happens to the monopoles as they collide: the results of this are
shown in Figure~\ref{fig4}.
In effect, the scattering takes place in the plane $x=0$, and so we
use contour plots in this plane. Note that the range of $z$ in the
plots has been shifted, using the periodicity, in order to give a clearer
representation of the scattering. As $W$ increases from a small positive
value, the monopoles come in along the $y$-axis, with $z=0$. By
$W=1/2$ (the first plot in Figure~\ref{fig4}) we see that they have merged at
the origin. Then (second plot, with $W=2/3$) they begin to separate
along the $z$-axis. At $W=1$, they are equidistant: this is a special
solution (\ref{SpecialPoints}). Then they re-merge at $z=\pi$ on the
$z$-axis, and separate in the $\pm y$-directions. The combination
of $90^{\circ}$ scattering in the $yz$-plane and periodicity in $z$ leads,
in this example, to a picture in which monopoles emerge in the same
directions as they entered, but shifted by half a period.
The plots in Figure~\ref{fig4} are for $C=1$; for other
values of $C$, the picture is qualitatively the same, although the
details differ --- for example, the value of $W$ at which the monopoles
merge.

Let us consider, next, the geodesic $W=\ii p^2$ with $p>0$,
{\sl ie.}\ $W$ on the positive imaginary axis.
If $p\gg1$, then $K\approx\ii p^2$ and the monopoles are located at
$x+\ii y=\pm Cp\ee^{-\ii\pi/4}$, $z=\pi$; while if $0<p\ll1$, then
$K\approx-\ii p^{-2}$ and the monopoles are located at
$x+\ii y=\pm Cp^{-1}\ee^{\ii\pi/4}$, $z=0$. So here the trajectory
is fully 3-dimensional: the monopoles undergo right-angle
scattering in the $xy$-direction, as well as being shifted by half
a period in $z$.

As long as $C$ is not too large, any radial line in the $W$-plane is an
approximate
geodesic representing head-on scattering of two monopoles, and it is
easy to see that we get a picture which interpolates between the two
examples above. In fact, the line $W=p^2\ee^{\ii\nu}$, with $\nu$ fixed,
gives a scattering angle of $\nu$, in addition to the $z$-shift.
For large $C$, however, one gets rather different trajectories:
see, for example, figure~6 of \cite{M13}.


\section{Concluding remarks}

We have seen that the periodic monopole system admits dynamical
behaviour not seen in the non-periodic $\RR^3$ case, in particular
head-on collision of two monopoles resulting in scattering through
any angle, accompanied by a half-period shift. This is a consequence
both of the periodicity, and of the absence of rotational symmetry
about the periodic axis. It would be worth studying other geodesics
on the double-trumpet geodesic surface, not just those representing
head-on collisions, as this might reveal further novel behaviour.

We have shown that the asymptotic metric on the 2-monopole moduli
space may be computed directly via a simple approximation of the
relevant Nahm/Hitchin data; and this metric agrees, as expected, with
that derived by considering the effective 2-monopole Lagrangian.
This asymptotic metric is relatively simple, having continuous
symmetries; in particular, one can identify geodesic
surfaces which are different from those described in this paper, and
which do not involve $\theta$ and $\omega$ remaining constant.
It would be interesting to investigate the global structure of these,
and to look for geodesics (trajectories) where $\theta$ and/or
$\omega$ change.

The methods used here should extend to the case of higher charge
monopoles. In the $\RR^3$ case, it is particularly useful to consider
multi-monopoles invariant under discrete subgroups of the rotation
group \cite{HMM95, MS04}. Because of the lack of rotational symmetry
in the periodic case, it seems unlikely that the full scope of this
technique could be applied. But certain discrete symmetries such as
cyclic symmetry should remain relevant, and it would be worth making
a systematic study of the symmetries of the SU($n$) Hitchin system,
corresponding to centred periodic $n$-monopole solutions, for higher
values of $n$. Some preliminary results along these lines have been
obtained, and further work is in progress.

\section*{Acknowledgments}
RM was supported by a studentship, and RSW by the Consolidated Grant
ST/J000426/1, both from the UK Science and Technology Facilities Council.

\end{document}